\documentclass[conference]{IEEEtran}
\IEEEoverridecommandlockouts
\usepackage{cite}
\usepackage{amsmath,amssymb,amsfonts}
\usepackage{algorithmic}
\usepackage{graphicx}
\usepackage{textcomp}
\usepackage{xcolor}
\usepackage{mathtools}
\usepackage{bbm}
\usepackage{amsmath}
\usepackage[short]{optidef}

\usepackage[left=0.625in,right=0.625in,top=0.75in,bottom=1.05in]{geometry}

\def\BibTeX{{\rm B\kern-.05em{\sc i\kern-.025em b}\kern-.08em
T\kern-.1667em\lower.7ex\hbox{E}\kern-.125emX}}

\title{Outage Probability Analysis of Tunable Liquid Lens-assisted VLC Systems\vspace{-5mm}\\}

\author{
\IEEEauthorblockA{\normalsize{Kapila W. S. Palitharathna}\IEEEauthorrefmark{1}, \normalsize{Nikos  A. Mitsiou}\IEEEauthorrefmark{1}, \normalsize{Constantinos Psomas}\IEEEauthorrefmark{2}, and \normalsize{Ioannis Krikidis}\IEEEauthorrefmark{1}}
\IEEEauthorblockA{\IEEEauthorrefmark{1}Department of Electrical and Computer Engineering, University of Cyprus, Nicosia, Cyprus\\}
\IEEEauthorblockA{\IEEEauthorrefmark{2}Department of Computer Science and Engineering, European University Cyprus, Nicosia, Cyprus\\}
\IEEEauthorblockA{Email: \{palitharathna.kapila, mitsiou.nikolaos, krikidis\}@ucy.ac.cy, c.psomas@euc.ac.cy\vspace{-10mm}}\thanks{This work received funding from the Smart Networks and Services Joint Undertaking (SNS JU) under the European Union's Horizon Europe research and innovation programme (Grant Agreement No. 101192080) and from the European Union’s Erasmus+ Programme under the CoRe 5G project (Grant Agreement No. 101260846). It was also funded by the SMALL SCALE INFRASTRUCTURES/1222/0087 LARISSA-6G under THALEIA 2021-2027.}\vspace{-4mm}}

\begin{document}

\maketitle

\begin{abstract}
This paper presents a tunable liquid lens (TLL)-assisted indoor mobile visible light communication system. To mitigate performance degradation caused by user mobility and random receiver orientation, an electrowetting cuboid TLL is used at the receiver. By dynamically controlling the orientation angle of the liquid surface through voltage adjustments, signal reception and overall system performance are enhanced. An accurate mathematical framework is developed to model channel gains, and two lens optimization strategies, namely ($i$) the best signal reception (BSR), and ($ii$) the vertically upward lens orientation (VULO) are introduced for improved performance. Closed form expressions for the outage probability are derived for each scheme for practical mobility and receiver orientation conditions. Numerical results demonstrate that the proposed TLL and lens adjustment strategies significantly reduce the outage probability compared to fixed lens and no lens receivers across various mobility and orientation conditions. Specifically, the outage probability is improved from $1\times 10^{-1}$ to 
$3\times 10^{-3}$ at a transmit power of 
$12$ dBW under an $8^{\circ}$ polar angle variation in random receiver orientation using the BSR scheme. 
\end{abstract}


\vspace{-4mm}
\section{Introduction}\label{introduction}
\vspace{-1mm}
Future wireless networks are expected to deliver ultra-high data rates, low latency, and enhanced security while supporting a wide range of emerging applications such as augmented reality, e-health, and Industry 4.0. To meet these demands, visible light communication (VLC) has emerged as a promising technology, particularly for indoor short-range communication. By utilizing the broad visible light spectrum (380 nm to 780 nm), VLC offers a significantly larger bandwidth compared to traditional radio frequency (RF) communication, enabling high data rate transmission~\cite{Ghassemlooy}. In addition to providing high-speed data transfer, VLC systems offer the dual functionality of energy-efficient illumination and wireless communication, leveraging low-cost light-emitting diodes (LEDs) as transmitters and photodiodes (PDs) as receivers, ensuring seamless integration with existing lighting infrastructure~\cite{Ghassemlooy}.

Despite these advantages, VLC systems rely heavily on a strong line-of-sight (LoS) channel for efficient data transmission. The availability of a reliable LoS link is influenced by several factors, including transmitter-receiver alignment, blockages from static objects/humans, random receiver orientation, and user mobility. These challenges can significantly degrade system performance, making it essential to develop robust solutions. Several studies have investigated the impact of these conditions on the VLC performance and proposed strategies for performance enhancement~\cite{Srivastava_2021, Beysens_2020, Abumarshoud_2022, Kapila_2022}. In~\cite{Srivastava_2021}, the downlink performance of an indoor VLC system has been analyzed under static/mobile blockage conditions. A user-guidance system to avoid blockages and enhance link reliability has been introduced in~\cite{Beysens_2020}. In~\cite{Abumarshoud_2022}, an intelligent reflecting surface (IRS)-assisted non-orthogonal multiple access VLC system has been proposed to improve link reliability. Existing research has also explored machine learning techniques for performance enhancement under random receiver orientation and user mobility conditions~\cite{Kapila_2022}.

Recently, optical lens technology has introduced several reconfigurable liquid lens architectures, offering a promising approach to enhance communication efficiency by dynamically optimizing the lens properties for improved signal reception~\cite{Ndjiongue_2021, Ngatched_20212, Tian_2022, Kapila_2025}. Among these technologies, liquid crystal (LC)-based structures, which adjust the refractive index to manipulate light propagation, have been explored in the context of VLC systems~\cite{Ndjiongue_2021, Ngatched_20212}. In~\cite{Ndjiongue_2021} and~\cite{Ngatched_20212}, LC-based IRSs were employed at the receiver to dynamically steer light beams toward the PD, enhancing overall system performance. In addition, mechanical liquid lenses have been investigated for their potential in dynamic beam steering \cite{Kapila_2025}. Studies in \cite{Kapila_2025} examined an adaptable liquid lens with three degrees of freedom i.e., focal length, azimuth angle, and polar angle, to enhance communication efficiency by mitigating inter-channel interference in multiple-input multiple-output VLC systems. However, they often rely on complex channel models that are mathematically intractable, requiring extensive simulations for performance evaluation. 

Although largely omitted in the context of VLC systems, several non-mechanical tunable liquid lens (TLL) architectures have been proposed that can change the orientation and shape of the liquid surface smoothly to control the direction of light propagation~\cite{Cheng_2021,Lee_2019,Zohrabi_2016}. In~\cite{Lee_2019}, an innovative three-dimensional beam steering methodology that leverages an electrowetting TLL in conjunction with a liquid prism has been introduced, enhancing the precision of light manipulation. In~\cite{Zohrabi_2016}, the authors presented techniques for one- and two-dimensional beam steering employing multiple TLLs, demonstrating significant advancements in beam control capabilities. However, to the best of the authors' knowledge this is the first study that uses such an TLL to improve the performance of VLC systems under random receiver orientation and user mobility, while analyzing the system's performance.

In this paper, we investigate an indoor VLC system consisting of a single access point (AP) and a mobile user equipped with a VLC receiver. The receiver incorporates an electrowetting surface-based cubic TLL capable of dynamically adjusting its liquid surface based on the user's position and receiver orientation to address practical challenges such as random receiver orientation and user mobility, thereby improving system performance. We present an accurate mathematical model to analyze the channel gain. To optimize signal reception, we introduce two TLL orientation schemes with varying levels of complexity, namely ($i$) the best signal reception (BSR), and ($ii$) the vertical upward lens orientation (VULO). The average outage probability for each scheme is analyzed, and closed-form expressions for the outage probability are derived. Numerical results validate the analytical framework and demonstrate the performance gains of the proposed TLL architecture. In particular, the BSR scheme achieves significant improvements in outage probability, compared to the receivers where no lens is deployed or a fixed lens is used. 

    \begin{figure}[!t]
        \centering
        \includegraphics[width=0.65\columnwidth]{"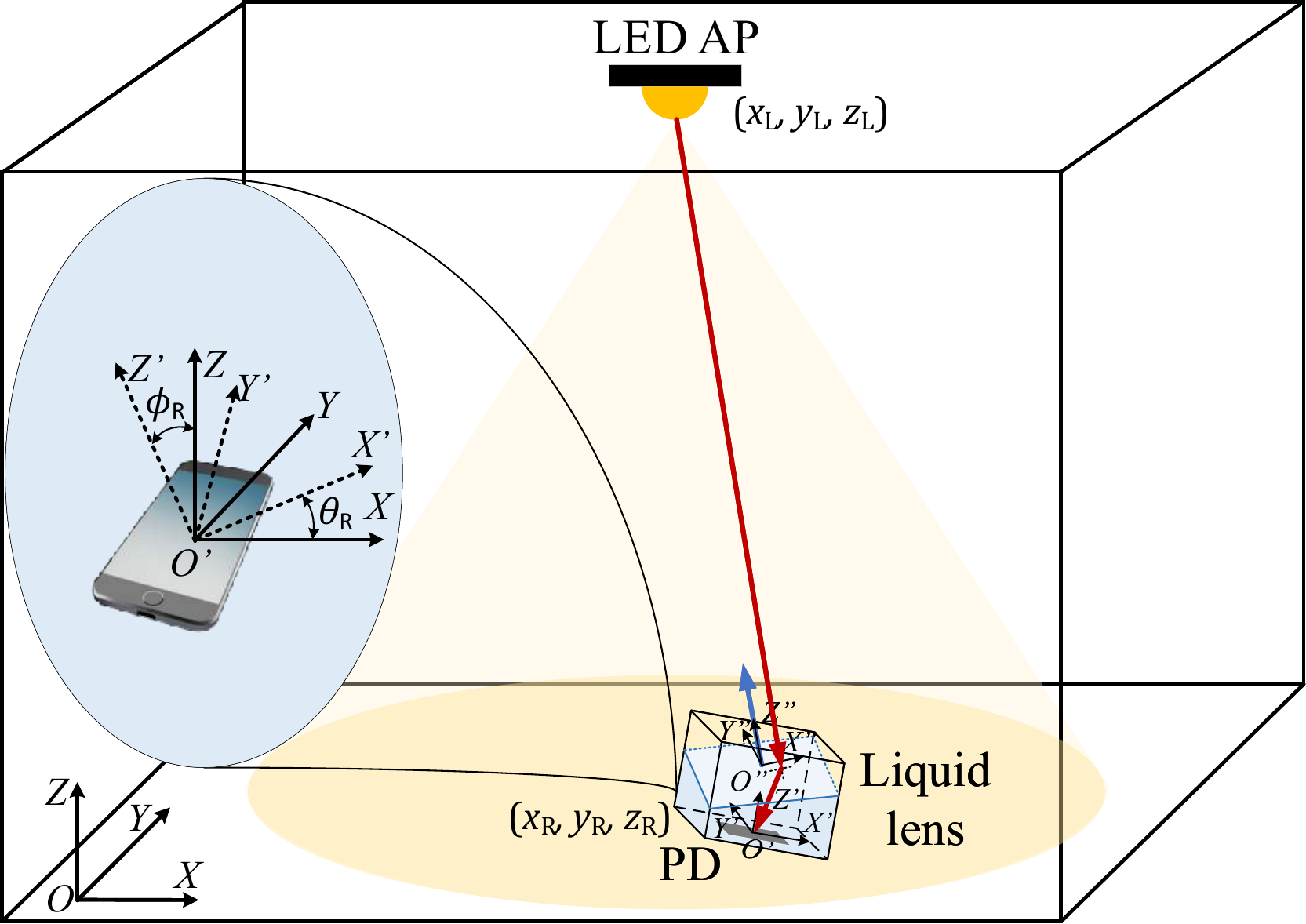"}\
        \vspace{-3mm}
        \caption{An indoor VLC system with a TLL-assisted receiver.}
        \label{fig:f1}
        \vspace{-7mm}
    \end{figure}

\vspace{-1mm}
\section{System Model}\label{system}
We consider an indoor VLC system comprising a ceiling-mounted LED that serves as the AP, and a mobile user equipped with a TLL-assisted receiver, as shown in Fig. \ref{fig:f1}. The AP is positioned at $\hat{\boldsymbol{P}}_{L}\hspace{-0.5mm} = \hspace{-0.5mm}[x_{L}, y_{L}, z_{L}]^T$ with a vertically downward orientation. The user device, subject to mobility and random receiver orientation, is positioned at $\hat{\boldsymbol{P}}_{R} \hspace{-0.5mm}=\hspace{-0.5mm} [x_{R}, y_{R}, z_{R}]^T$. For modeling purposes, we define the room's coordinate frame, the receiver's local coordinate frame, and the lens surface's coordinate frame as $OXYZ$, $O'X'Y'Z'$, and $O''X''Y''Z''$, respectively. The receiver's orientation is characterized by an azimuth angle $\theta_{R}\sim \mathcal{U}(0, 2\pi)$ and a polar angle $\phi_{R}\sim \mathcal{N}(\mu_{\phi_R}, \sigma_{\phi_R}^2)$, which are modeled to align with empirical measurements of mobile users~\cite{soltani_2019}. To model user mobility, we adopt a 2D topology of the random waypoint (RWP) model, where the velocity and destination points (waypoints) are randomly selected within a circular radius of $R_{circ}$~\cite{Arfaoui_2021}. 

\vspace{-1mm}
\subsection{TLL Model}\label{liquid lens}
\begin{figure}[!t]
        \centering
        \includegraphics[width=0.6\columnwidth]{"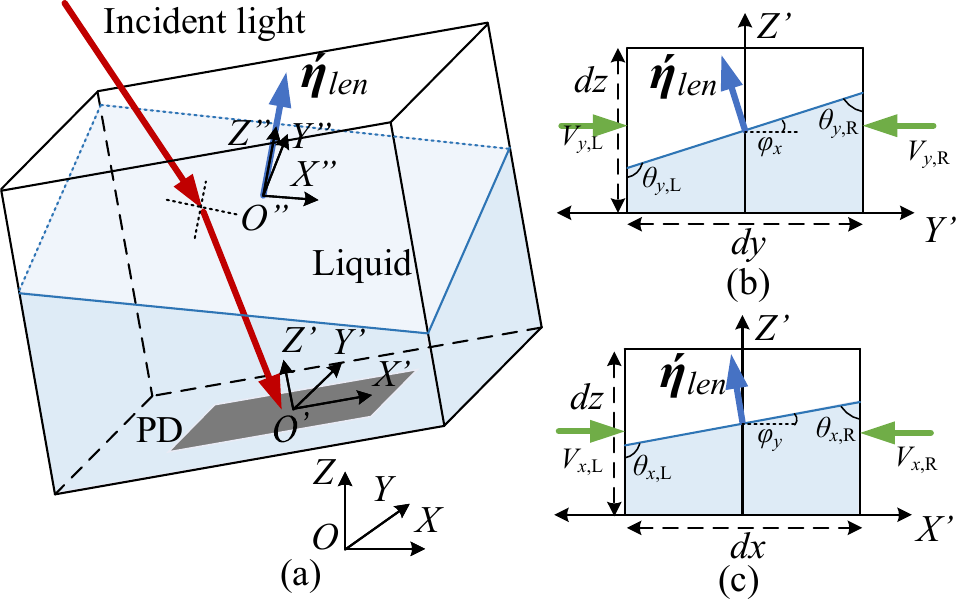"}\
        \vspace{-3mm}
        \caption{Electrowetting TLL-assisted VLC receiver. (a) Perspective view. (b) View from the $X'$ direction. (c) View from the $Y'$ direction.}
        \label{fig:f2}
        \vspace{-7mm}
    \end{figure}

As illustrated in Fig. \ref{fig:f2}(a), the proposed electrowetting TLL is a cuboid structure with dimensions $d_x\times d_y \times d_z$, partially filled with an optically transparent liquid. The four side surfaces of the lens are composed of electrowetting surfaces~\cite{Lee_2013,Cheng_2021}. By varying the voltages applied to these surfaces, namely $V_{x,L}$, $V_{x,R}$, $V_{y,L}$, and $V_{y,R}$, the corresponding contact angles at the walls, $\theta_{x,L}$, $\theta_{x,R}$, $\theta_{y,L}$, and $\theta_{y,R}$, can be dynamically adjusted~\cite{Lee_2013}. This variation in contact angles modifies the normal vector of the liquid surface, denoted as $\hat{\boldsymbol{\eta}}_{len}$, which can be controlled through the applied voltages. Fig. \ref{fig:f2}(b) and Fig. \ref{fig:f2}(c) present side views of the TLL from the $X'$ and $Y'$ directions, respectively. Assuming an initial contact angle of $90^{\circ}$, the relationship between the applied voltages and the contact angle can be expressed as~\cite{Lee_2013}

    \vspace{-6mm}
    \begin{align}\label{equ:e1}
        \hspace{-2mm}\theta_{M,N} = \cos^{-1}{\bigg(\frac{kV_{M,N}^2}{d_{M}}\bigg)}, \forall M\in\left\{x,y\right\}, \forall  N\in \left\{L,R\right\},
        \vspace{-4mm}
    \end{align}
where $k = \frac{\varepsilon_0\varepsilon_1}{2\gamma_{LV}}$, with $\varepsilon_1$ representing the relative permittivity of the dielectric layer, 
$\varepsilon_0$ denoting the permittivity of vacuum, and $\gamma_{LV}$ being the interfacial surface tension between the liquid and vapor phases~\cite{Cheng_2021}. Furthermore, under the assumption that the liquid surface remains flat, the relationship between the voltages applied to two opposing electrowetting surfaces can be expressed as
    \vspace{-3mm}
    \begin{align}\label{equ:e2}
        \hspace{-2mm}\sum_{ M\in\left\{x,y\right\}}\cos^{-1}\bigg(\frac{kV_{M,N}^2}{d_{M}}\bigg) = \pi, \quad \quad \forall N\in \left\{L,R\right\}.
    \vspace{-4mm}
    \end{align}
Using basic trigonometry, the angles of the liquid surfaces with respect to (w.r.t.) the $X'Y'$ plane on the $X'Z'$ and $Y'Z'$ planes can be expressed as $\psi_{M} = \frac{\pi}{2}-\theta_{M,R}, \forall M$. 

\vspace{-2mm}
\section{VLC Channel Model}\label{channel}
The optical channel gain $h_{L,R}$ from the LED to the PD is expressed using the well-known Lambertian model as~\cite{soltani_2019}
    \vspace{-2mm}
    \begin{equation}\label{equ:e10}
		h_{L,R}	=
		\displaystyle\frac{(m+1)A}{2\pi d_{L,R}^2}\cos^m(\theta_{L,R})\cos(\phi_{L,R})\Pi\left(\frac{\phi_{L,R}}{\phi_{FoV}}\right),
        \vspace{-2mm}
	\end{equation}
where $m=-\ln(2)/\ln(\cos(\theta_{1/2}))$ represents the Lambertian order of the LED, $A$ is the aperture area of the PD, $d_{L,R}$ is the Euclidean distance between the LED and the receiver, $\theta_{1/2}$ is the half-power semi-angle of the LED, $\theta_{L,R}$ is the irradiance angle of the LED, $\phi_{L,R}$ is the incident angle at the PD, $\phi_{FoV}$ is the field-of-view (FoV) of the PD, and $\Pi(x)$ is the rectangular function, equal to 1 for $|x|\le 1$ and $0$ otherwise.
\vspace{-2mm}
\subsection{Refraction of Light from the TLL}
We derive analytical expressions for the channel gain, $h_{L,R}$ of the VLC link incorporating the proposed TLL. To achieve this, we compute the rotation matrices, unit normal vectors, and the direction of light refraction. The receiver's coordinate frame, $O'X'Y'Z'$, is modeled as a rotated version of the room's coordinate frame, $OXYZ$, obtained by a rotation of $\theta_{R}$ around the $z$-axis followed by a rotation of $\phi_{R}$ around the $y$-axis. The combined rotation is represented by the rotation matrix $^0\mathbf{R}_1(\theta_{R},\phi_{R})=\mathbf{R}_y(\phi_{R})\mathbf{R}_z(\theta_{R})$, where $\mathbf{R}_y(\phi_{R})$ and $\mathbf{R}_z(\theta_{R})$ denote the rotation matrices about the $y$-axis and $z$-axis, respectively. The transformation simplifies to
    \vspace{-2mm}
    \begin{equation}\label{equ:rot01}
    \vspace{2mm}
        \hspace{-2.5mm}^0\mathbf{R}_1(\theta_{R},\phi_{R})
        = 
        \begin{bmatrix}
         \text{c}{\theta_{R}}\text{c}{\phi_{R}} & \text{s}{\theta_{R}}\text{c}{\phi_{R}} & -\text{s}{\phi_{R}}\\
         -\text{s}{\theta_{R}} & \text{c}{\theta_{R}}& 0\\
        \text{c}{\theta_{R}}\text{s}{\phi_{R}} & \text{s}{\theta_{R}}\text{s}{\phi_{R}} & \text{c}{\phi_{R}}\\
        \end{bmatrix},
        \vspace{-1mm}
    \end{equation}
where $\cos{\theta}$, and $\sin{\theta}$ are denoted as $\text{c}\theta$ and $\text{s}\theta$, respectively. The normal vector to the receiver is aligned along the $Z'$ axis. Therefore, using the passive rotation transformation between the two coordinate frames, the unit normal vector of the receiver in the room's coordinate frame is given by $\hat{\boldsymbol{\eta}}_{R}={^0\mathbf{R}_1^{-1}(\theta_{R},\phi_{R})[0\quad 0 \quad 1]^T}$. Expanding this expression, $\hat{\boldsymbol{\eta}}_{R}$ can be expressed as
\vspace{-2mm}
    \begin{equation}\label{equ:rot01}
        \hat{\boldsymbol{\eta}}_{R}
        = [
        \text{c}{\theta_{R}}\text{s}{\phi_{R}},\quad 
        \text{s}{\theta_{R}}\text{s}{\phi_{R}},\quad
        \text{c}{\phi_{R}}]^T.
        \vspace{0mm}
    \end{equation}

The transformation between the receiver's coordinate frame $O'X'Y'Z'$ and the coordinate frame at the liquid surface $O''X''Y''Z''$ is represented by the rotation matrix $^1\mathbf{R}_2(\psi_{x},\psi_{y})=\mathbf{R}_y(\psi_{y})\mathbf{R}_x(\psi_{x})$ where $\mathbf{R}_y(\psi_{y})$ and $\mathbf{R}_x(\psi_{x})$ denote rotations about $y$- and $x$-axis, respectively. Expanding these transformations, the rotation matrix simplifies to
    \vspace{-1mm}
    \begin{equation}\label{equ:rot01}
        \hspace{-2.5mm}^1\mathbf{R}_2(\psi_{x},\psi_{y})
        = 
        \begin{bmatrix}
         \text{c}{\psi_{y}} & \text{s}{\psi_{y}}\text{s}{\psi_{x}} & \text{s}{\psi_{y}}\text{c}{\psi_{x}}\\
         0 & \text{c}{\psi_{x}}& -\text{s}{\psi_{x}}\\
        -\text{s}{\psi_{y}} & \text{c}{\psi_{y}}\text{s}{\psi_{x}} & \text{c}{\psi_{y}}\text{c}{\psi_{x}}\\
        \end{bmatrix}.
        \vspace{-1mm}
    \end{equation}
The unit normal vector to the lens is aligned along the $z''$-axis in its own coordinate frame. Expressing this vector in the room’s coordinate frame, we have $\hat{\boldsymbol{\eta}}_{len}= {^0\mathbf{R}_1^{-1}(\theta_{R},\phi_{R})}^1\mathbf{R}_2^{-1}(\psi_{x},\psi_{y})[0\quad 0 \quad 1]^T$. By expanding and simplifying the transformations, $\hat{\boldsymbol{\eta}}_{len}$ is given by
    \vspace{-1mm}
    \begin{equation}\label{eta_len1}
        \hspace{-1mm}\hat{\boldsymbol{\eta}}_{len}
        = 
        \begin{bmatrix}
         \text{c}{\theta_{R}}\text{s}{\phi_{R}}\text{c}{\psi_{y}}\text{c}{\psi_{x}}-\text{s}{\theta_{R}}\text{c}{\psi_{y}}\text{s}{\psi_{x}}-\text{c}{\theta_{R}}\text{c}{\phi_{R}}\text{s}{\psi_{y}}\\
         \text{s}{\theta_{R}}\text{s}{\phi_{R}}\text{c}{\psi_{y}}\text{c}{\psi_{x}}+\text{c}{\theta_{R}}\text{c}{\psi_{y}}\text{s}{\psi_{x}}-\text{s}{\theta_{R}}\text{c}{\phi_{R}}\text{s}{\psi_{y}}\\
         \text{c}{\phi_{R}}\text{c}{\psi_{y}}\text{c}{\psi_{x}}+\text{s}{\phi_{R}}\text{s}{\psi_{y}}\\
        \end{bmatrix}.
        \vspace{0mm}
    \end{equation}

The direction vector along the receiver-transmitter trajectory is denoted as $\hat{\boldsymbol{\eta}}_{TR} = [e_{TR,x}, e_{TR,y}, e_{TR,z}]^T = \frac{\hat{\boldsymbol{P}}_{L}-\hat{\boldsymbol{P}}_{R}}{||\hat{\boldsymbol{P}}_{L}-\hat{\boldsymbol{P}}_{R}||}$. The light refraction direction vector, $\hat{\boldsymbol{\eta}}_{ref}$, can be derived using the vector form of Snell's law, and can be expressed as~\cite{Kapila_2025}
    \vspace{-1mm}
    \begin{align}\label{equ:ref_o}
        \hspace{-2mm}\hat{\boldsymbol{\eta}}_{ref} &= n_l^{-1}\big[\hat{\boldsymbol{\eta}}_{len}\times (\hat{\boldsymbol{\eta}}_{len}\times \hat{\boldsymbol{\eta}}_{TR}) \nonumber\\
        &- \hat{\boldsymbol{\eta}}_{len}\left(n_l^2-(\hat{\boldsymbol{\eta}}_{len}\times\hat{\boldsymbol{\eta}}_{TR})\cdot(\hat{\boldsymbol{\eta}}_{len}\times\hat{\boldsymbol{\eta}}_{TR})\right)^{-\frac{1}{2}}\big],\nonumber \\
        &=\hspace{-0.5mm}{n_{l}}^{-1}\big(\hat{\boldsymbol{\eta}}_{len}\big(\text{c}(\alpha_{T,R})-(n_l^2 \hspace{-0.5mm}- \hspace{-0.5mm}\text{s}^2(\alpha_{T,R}))^{\frac{1}{2}}\big)\hspace{-0.5mm}-\hspace{-0.5mm}\hat{\boldsymbol{\eta}}_{TR}\big),
        \vspace{-3mm}
    \end{align}
where $n_l$ is the relative refractive index of the liquid. This simplification was obtained utilizing the vector identity, $\boldsymbol{A}\times (\boldsymbol{A}\times \boldsymbol{B}) = (\boldsymbol{A}\cdot\boldsymbol{B})\cdot\boldsymbol{A} - (\boldsymbol{A}\cdot\boldsymbol{A})\cdot \boldsymbol{B}$ along with the relation $\text{s}^2(\alpha_{T,R}) =(\hat{\boldsymbol{\eta}}_{len}\times\hat{\boldsymbol{\eta}}_{TR})\cdot(\hat{\boldsymbol{\eta}}_{len}\times\hat{\boldsymbol{\eta}}_{TR})$, where $\alpha_{T,R} = \cos^{-1}{\left(\hat{\boldsymbol{\eta}}_{len}\cdot\hat{\boldsymbol{\eta}}_{TR}\right)}$ is the light incident angle at the liquid surface. 

\vspace{-2mm}
\subsection{Simplified Channel Model}
We derive a simplified expression for $h_{L,R}$ as a function of the receiver position, receiver orientation, and the orientation of the lens surface. The $h_{L,R}$ in~\eqref{equ:e10} can be re-written as 
     \vspace{-1mm}
    \begin{align}\label{channel_sim}
        h_{L,R} = k_1||\hat{\boldsymbol{P}}_{L}-\hat{\boldsymbol{P}}_{R}||^{-(m+2)} \text{c}\left(\phi_{L,R}\right),
         \vspace{-2mm}
    \end{align}
where $k_1 = \frac{(m+1)A(z_L-z_R)^m}{2\pi}$ is a constant, the term $||\hat{\boldsymbol{P}}_{L}-\hat{\boldsymbol{P}}_{R}||^{-(m+2)}$ remains invariant for a given user position, and $\phi_{L,R}$ represents the light incident angle at the PD after the reflection from the liquid surface. The angle $\phi_{L,R}$ is the angle between the unit vectors $-\hat{\boldsymbol{\eta}}_{ref}$ and $\hat{\boldsymbol{\eta}}_{R}$. Consequently, $\text{c}(\phi_{L,R})$ can be expressed as
    \vspace{-2mm}
    \begin{align}
        \text{c}(\phi_{L,R}) = -\hat{\boldsymbol{\eta}}_{ref}\cdot\hat{\boldsymbol{\eta}}_{R}.
        \vspace{-2mm}
    \end{align}
To determine $\text{c}(\phi_{L,R})$, we take the dot product of equation~\eqref{equ:ref_o} with 
$-\hat{\boldsymbol{\eta}}_{R}$, yielding
    \vspace{-1mm}
    \begin{align}\label{cos_theta}
        \text{c}(\phi_{L,R}) \hspace{-0.5mm}= \hspace{-0.5mm}-{n_{l}}^{-1}\big(\big(\text{c}(\alpha_{T,R})-(n_l^2 \hspace{-0.5mm}&- \hspace{-0.5mm}\text{s}^2(\alpha_{T,R}))^{\frac{1}{2}}\big)(\hat{\boldsymbol{\eta}}_{len}\cdot \hat{\boldsymbol{\eta}}_{R}\big)\hspace{-0.5mm}\nonumber \\
        &-\hspace{-0.5mm}\big(\hat{\boldsymbol{\eta}}_{TR}\cdot\hat{\boldsymbol{\eta}}_{R}\big)\big).
    \end{align}
We observe that adjusting the values of $\psi_x$ and $\psi_y$ to maximize $\text{c}(\phi_{L,R})$ in \eqref{cos_theta} can help mitigate the performance degradation caused by random receiver orientation of the receiver. 

\vspace{-1mm}
\section{Lens Orientation Schemes}\label{lens_orientation}
We propose two lens orientation schemes, namely ($i$) BSR and ($ii$) VULO, which enhance signal reception by optimizing the values of $\psi_x$ and $\psi_y$ and compare them with two conventional receiver configurations: one with a fixed lens and one without a lens. 

\subsection{BSR Scheme}
In the BSR scheme, we adjust the lens orientation to maximize the channel gain of the VLC link from the AP to the receiver. It is observed that the channel gain, denoted as $h_{L,R}$, reaches its maximum value when $\text{c}(\phi_{L,R}) = 1$ in \eqref{channel_sim}. To determine the corresponding lens orientation angles, $\text{c}\psi_x$ and $\text{c}\psi_y$, the following steps are used. First, we set $\text{c}(\phi_{L,R}) = 1$ in \eqref{cos_theta}. By expanding $\hat{\boldsymbol{\eta}}_{len}\hspace{-1.5mm}\cdot\hat{\boldsymbol{\eta}}_{R}$ and using the fundamental trigonometric identity $\text{c}^2{\theta}+\text{s}^2{\theta} = 1$, we can express $\hat{\boldsymbol{\eta}}_{len}\hspace{-1.5mm}\cdot\hat{\boldsymbol{\eta}}_{R}$ as $\text{c}\psi_x\text{c}\psi_y$. Consequently, in the BSR scheme, the expression in \eqref{cos_theta} can be simplified to 
\vspace{-2mm}
\begin{align}\label{cos_theta_1}
        n_l\hspace{-0.5mm}- \hspace{-0.5mm}\hat{\boldsymbol{\eta}}_{TR}\cdot \hat{\boldsymbol{\eta}}_{R}\hspace{-0.5mm}+\hspace{-0.5mm} \left(\text{c}(\alpha_{T,R})-(n_l^2 - \text{s}^2(\alpha_{T,R}))^{\frac{1}{2}}\right)\text{c}\psi_x\text{c}\psi_y=0.
    \end{align}
We use numerical methods to solve \eqref{cos_theta_1} and obtain the optimal values of $\psi_{x}$ and $\psi_{y}$, respectively. A random value for $\psi_x\in[\psi_x^{L},\psi_x^{H}]$ is selected, and~\eqref{cos_theta_1} is solved to obtain $\psi_y\in[\psi_y^{L},\psi_y^{H}]$. This is repeated to obtain a feasible solution.

\subsection{VULO Scheme}
In the VULO scheme, orientation of the lens surface is kept vertically upward, regardless of the receiver orientation and the position. This approach is considered as a low complexity scheme compared to the BSR scheme. In this scheme, $\hat{\boldsymbol{\eta}}_{len} = 
[0, 0, 1]^T$, which results in $\text{c}(\alpha_{L,R}) = e_{TR,z}$. Additionally, $\hat{\boldsymbol{\eta}}_{len}\cdot\hat{\boldsymbol{\eta}}_{R} = \text{c}(\phi_{R})$. As a result, the expression for $\text{c}(\phi_{L,R})$ in \eqref{cos_theta} can be further simplified as 
    \vspace{-2mm}
    \begin{align}\label{cos_theta_3}
        \text{c}(\phi_{L,R}) = -{n_{l}^{-1}}\bigg(e_{TR,z}-(n_l^2 &- (1-e_{TR,z}^2)))^{-\frac{1}{2}}\bigg)\text{c}(\phi_{R})\nonumber \\
        &+n_{l}^{-1}\left(\hat{\boldsymbol{\eta}}_{TR}\cdot \hat{\boldsymbol{\eta}}_{R}\right).
        \vspace{-5mm}
    \end{align}
To determine the optimal values of $\psi_{x}$ and $\psi_{y}$, the following steps are employed. In this scheme, the components of $\hat{\boldsymbol{\eta}}_{len}$ along the $x$- and $y$-axes are set to zero. By multiplying the $x$-axis component by $\text{s}\theta_R$ and the $y$-axis component by $\text{c}\theta_R$ in \eqref{eta_len1}, and subsequently summing the resulting expressions, we derive
    \vspace{-5mm}
    \begin{align}\label{equ_122}
        \text{s}{\phi_{R}}\text{c}{\psi_{y}}\text{c}{\psi_{x}}-\text{c}{\phi_{R}}\text{s}{\psi_{y}}=0.
        \vspace{-2mm}
    \end{align}
Next, we combine \eqref{equ_122} with the component along the $z$-axis of $\hat{\boldsymbol{\eta}}_{len}$ in \eqref{eta_len1} to solve for the optimal $\psi_{y}$. In the VULO scheme, the optimal $\psi_{y}$ is given by $\phi_{R}$. By substituting $\psi_{y} = \phi_{R}$ into \eqref{equ_122}, we obtain the optimal $\psi_{x}=0^{\circ}$.

\section{Outage Probability Analysis}
We present the expressions for the outage probability of the proposed system and derive closed-form exact/approximate expressions for the outage probability of each lens orientation scheme. The outage probability of the proposed system can be written as 
    \vspace{-2mm}
    \begin{align}\label{prob_out}
        P_O(\mu_{\phi_R},\sigma_{\phi_R}^2) = \mathbb{P}[{\rm SNR}(\theta_R,\phi_R,\mathbf{x})<\gamma_{th}], 
        \vspace{-2mm}
    \end{align}
where ${\rm SNR}(\theta_R,\phi_R,\mathbf{x})$ is signal-to-noise ratio (SNR) for a given receiver orientation angles $\theta_R$ 
and $\phi_R$, and user position $\mathbf{x}$. The parameter $\gamma_{th}$ denotes the threshold SNR value. The SNR at the receiver is ${\rm SNR}(\theta_R,\phi_R,\mathbf{x}) \hspace{-0.7mm}= \hspace{-0.7mm}R^2P_S^2h_{L,R}^2\hspace{-0.5mm}/\hspace{-0.5mm}\sigma^2\hspace{-1.5mm}$, where $R$ is the responsivity of the PD, $P_S$ is the transmit power at the LED, and $\sigma^2\hspace{-1mm}$ is the variance of the zero-mean additive white Gaussian noise. With the help of~\eqref{channel_sim} and using polar coordinates for the receiver position, the SNR can be rewritten as
    \vspace{-1mm}
    \begin{align}\label{SNR_gen}
        {\rm SNR}(\theta_R,\phi_R,\mathbf{x}) = \gamma_0 (r^2+h^2)^{-(m+2)}\text{c}^2(\phi_{L,R}),
    \end{align}
where $\gamma_0 = (P_S^2R^2(m+1)^2A^2h^{2m})/(4\pi^2\sigma^2)$. $r = ((x_R-x_L)^2+(y_R-y_L)^2)^{\frac{1}{2}}$ is the polar distance between the AP and the receiver, and $r\in [0, R_{circ}]$.

\subsection{Outage Probability of BSR Scheme}
In the BSR scheme, the light incident angle  $\phi_{L,R}$ is zero, as the light coming from the LED is perfectly aligned with the PD. As a result, the SNR expression in~\eqref{SNR_gen} simplifies to ${\rm SNR}(\theta_R,\phi_R) = \gamma_0(r^2+h^2)^{-(m+2)}$. To determine the outage probability in this scheme, we evaluate the probability density function (pdf) of the random variable (RV) $Y = \gamma_0(r^2+h^2)^{-(m+2)}$, denoted as $f_{Y}(y)$. The outage probability is then given by
    \vspace{-3mm}
    \begin{align}\label{prob_out8}
        P_O(\mu_{\phi_R},\sigma_{\phi_R}^2) \hspace{-0.5mm} &= \int_{0}^{\gamma_{th}} f_Y(y) dy.
        \vspace{-2mm}
    \end{align}

It is important to note that the RV $Y$ is independent of the receiver orientation angles $\theta_R$ and $\phi_R$, and depends solely on the distribution of the radial distribution $r$. The spatial user distributions for the RWP mobility model are polynomial functions of the distance, and its pdf is given by
\vspace{-2mm}
\begin{align}
    f_r(r) = \sum_{k=1}^n a_k\frac{r^k}{R_{circ}^{k+1}}, \quad 0\le r\le R_{circ},
    \vspace{-4mm}
\end{align}
where $n=3$, and the coefficients $[a_1, a_2, a_3] = \frac{1}{73}[324, -420, 96]$ are derived for a 2D topology as specified in~\cite{Arfaoui_2021}. For simplicity, we assume the pause time at the destination point is set to zero. To determine the pdf of $Y$, we apply the variable transformation $y= \gamma_0(r^2+h^2)^{-(m+2)}$ to the original pdf $f_r(r)$ using the relation $f_Y(y) = f_r(g^{-1}(y))\left|\frac{d(g^{-1}(y))}{dy}\right|$, and resulting pdf of $Y$ can be expressed as
\vspace{-2mm}
    \begin{align}\label{equ:pdf_Y}
        \hspace{-3mm} f_Y(y) = 
        \begin{cases}
            a_1y^{-b_{1}}+a_2y^{-b_{2}}+a_3y^{-b_{3}}, &  y_1\le y\le y_2,\\
            0, & \text{otherwise},
            \vspace{-2mm}
        \end{cases}
        \vspace{-8mm}
    \end{align}
where $a_1 = \frac{6\gamma_0^{m'}}{73(m+2)R_{circ}^2}\left(27+\frac{35h^2}{R_{circ}^2}+\frac{8h^4}{R_{circ}^4}\right)$, $a_2 = -\frac{6\gamma_0^{2m'}}{73(m+2)R_{circ}^4}\left(35+\frac{16h^2}{R_{circ}^2}\right)$, $a_3 = \frac{48\gamma_0^{3m'}}{73(m+2)R_{circ}^6}$, $b_1 = \frac{m+3}{m+2}$, $b_2 = \frac{m+4}{m+2}$, $b_3 = \frac{m+5}{m+2}$, $y_1 = \gamma_0(R_{circ}^2+h^2)^{-(m+2)}$, $y_2 = \gamma_0h^{-2(m+2)}$, and $m' = 1/(m+2)$. 

The integral in~\eqref{prob_out8} can be evaluated under three cases as follows. $\textit{Case 1: }$ $\gamma_{th} < y_1$. In this case, since $f_Y(y) = 0$ for $y<y_1$, the outage probability is simply $P_O(\mu_{\phi_R},\sigma_{\phi_R}^2) = 0$. $\textit{Case 2:}$ $y_1 < \gamma_{th} < y_2$. For this case, the outage probability can be given by the integral $P_O(\mu_{\phi_R},\sigma_{\phi_R}^2) = \int_{y_1}^{\gamma_{th}} f_Y(y) dy$. Substituting the expression for $f_{Y}(y)$ and performing the integration, the outage probability becomes
\vspace{-2mm}
\begin{align}
\vspace{-2mm}
    &\hspace{-2mm}P_O(\mu_{\phi_R},\sigma_{\phi_R}^2) = (m+2)\bigg[a_1\bigg(y_1^{-m'}-\gamma_{th}^{-m'}\bigg)\nonumber \\
    &\hspace{-2mm}+\frac{a_2}{2}\bigg(y_1^{-2m'}-\gamma_{th}^{-2m'}\bigg)+\frac{a_3}{3}\bigg(y_1^{-3m'}-\gamma_{th}^{-3m'}\bigg)\bigg].
    \vspace{-3mm}
\end{align}

$\textit{Case 3:}$ $y_2 < \gamma_{th}$. $P_O(\mu_{\phi_R},\sigma_{\phi_R}^2) = \int_{y_1}^{y_2} f_Y(y) dy$. Using the expression and with simple integrations the expression for $P_O(\mu_{\phi_R},\sigma_{\phi_R}^2)$ in this case reduces to 
\vspace{-2mm}
\begin{align}
\vspace{-4mm}
    \hspace{-1mm}&P_O(\mu_{\phi_R},\sigma_{\phi_R}^2)  = R_{circ}^2(m+2)\bigg[a_1\gamma_0^{-m'}+\frac{a_2}{2}\gamma_0^{-2m'}\nonumber \\
    &(R_{circ}^2+2h^2)\hspace{-0.5mm}+\hspace{-0.5mm}
    \frac{a_3}{3}\gamma_0^{-3m'}(3h^4\hspace{-0.5mm}+\hspace{-0.5mm}R_{circ}^4\hspace{-0.5mm}+\hspace{-0.5mm}3R_{circ}^2h^2)\bigg].
    \vspace{-4mm}
\end{align}
\vspace{-2mm}
    \begin{figure*}[!]
        \begin{align}\label{equ:out_VULO_111}
            f_{SNR(\mu_{\phi_R},\sigma_{\phi_R}^2)}(z)\hspace{-0.5mm}\approx \hspace{-0.5mm}{\frac{1}{\pi^2}}\hspace{-1.5mm}\sum_{k,i}\sum_{k',i'} C_{k,i}C'_{k',i'}B(1-b_k,1-b'_{k'})\sigma_{Y_{3}}^{1-b_k-b'_{k'}}G_{1,0}^{0,1}\left(-\frac{z^2}{2\sigma_{Y_3}^2}\middle\vert \frac{b_k+b'_{k'}-2}{2}\right).
            \tag{28}
        \end{align}
        \vspace{-7mm}
        \begin{align}\label{equ:out_VULO_112}
            P_{O}(\mu_{\phi_R},\sigma_{\phi_R}^2)\hspace{-0.5mm}\approx \hspace{-0.5mm}{\frac{1}{\pi^2}}\hspace{-1.5mm}\sum_{k,i}\sum_{k',i'} C_{k,i}C'_{k',i'}B(1-b_k,1-b'_{k'})\sigma_{Y_{3}}^{1-b_k-b'_{k'}}\sum_{j=1}^{n}w^1_jG_{1,0}^{0,1}\left(\frac{\gamma_{th}(1+x_j)}{2}\middle\vert \frac{b_k+b'_{k'}-2}{2}\right). \tag{29}
        \end{align}
        \vspace{-2mm}
        \hrulefill
        \vspace{-5mm}
    \end{figure*}
    
    \vspace{-4mm}
\subsection{Outage Probability of VULO Scheme}
In this scheme, the expression for $\text{c}\phi_{L,R}$ given in \eqref{cos_theta_3} can be simplified by using polar coordinates and applying standard trigonometric identities. Then, $\text{c}\phi_{L,R}$ can be expressed as 
\vspace{-3mm}
\begin{align}\label{equ:cosphi_VULO1}
    \text{c}\phi_{L,R} &= \left(1-r^2n_l^{-2}(r^2+h^2)^{-1}\right)^{-\frac{1}{2}}\text{c}\phi_R\nonumber \\
    &-n_l^{-1}r(r^2+h^2)^{-\frac{1}{2}}\text{s}\phi_R\text{c}(\theta-\theta_R).\tag{22}
    \vspace{-2mm}
\end{align}
By substituting the expression for $\text{c}\phi_{L,R}$ from \eqref{equ:cosphi_VULO1} into the general SNR formula in~\eqref{SNR_gen}, and subsequently expanding the resulting expression, we employ a Taylor series expansion to simplify the square root term in the expression. Through algebraic rearrangement—specifically substituting terms of the form $r^2(r^2+h^2)^{-(m+3)}$ with terms $(r^2+h^2)^{-(m+2)}-h^2(r^2+h^2)^{-(m+3)}$, the SNR in this scheme can be expressed as 
\vspace{-4mm}
    \begin{align}\label{SNR_VULO2}
        &{\rm SNR}(\theta_R,\phi_R,\mathbf{x}) \hspace{-0.5mm}= \hspace{-0.5mm}\sum_{k=1}^2(r^2\hspace{-0.5mm}+\hspace{-0.5mm}h^2)^{-(m+k+1)}\hspace{-0.5mm}(a_{k,1}\text{c}^2\phi_R\hspace{-0.5mm}\nonumber \\
        \vspace{-5mm}&+\hspace{-0.5mm}a_{k,2}\text{s}^2\phi_R\text{c}^2(\theta\hspace{-0.5mm}-\hspace{-0.5mm}\theta_R))\hspace{-0.5mm}+\hspace{-0.5mm}\hspace{-0.5mm} \sum_{k=0}^{\infty}a_{k,3}\binom{\frac{1}{2}}{k}r^{2k+1}(r^2\hspace{-0.5mm}+\hspace{-0.5mm}h^2)^{(k-m-\frac{5}{2})}\hspace{-0.5mm}\nonumber \\
        &\times\text{c}(\theta-\theta_R)\text{s}(2\phi_R),\tag{23}
        \vspace{-5mm}
    \end{align}
where $a_{1,1} = \gamma_0(1-n_l^{-2})$, $a_{1,2} = \gamma_0n_l^{-2}$, $a_{2,1} = \gamma_0h^2n_l^{-1}$, $a_{2,2} = -\gamma_0h^2n_l^{-2}$, and $a_{k,3}=\gamma_0n_l^{-(2k+1)}(-1)^k/2$.

To evaluate the pdf of the SNR, we first analyze the pdfs of the RVs constituting each term in \eqref{SNR_VULO2}. Consider the term $Y_1 = (r^2+h^2)^{-(m+2)}(a_{1,1}\text{c}^2\phi_R+a_{1,2}\text{s}^2\phi_R\text{c}^2(\theta-\theta_R))$, which can be observed as a product of two RVs, $Y_{1,1}=(r^2+h^2)^{-(m+2)}$ and $X_{1,1} = (a_{1,1}\text{c}^2\phi_R+a_{1,2}\text{s}^2\phi_R\text{c}^2(\theta-\theta_R))$. Assuming that both $\theta$, and $\theta_R$ are uniformly distributed over the interval $(0,2\pi)$, and using affine re-scaling, $\text{c}^2(\theta-\theta_R)\sim \text{Arcsine}(0,1)$. By applying variable transformation, we derive the conditional pdf of $X_{1,1}$ for a given $\phi_R$. Then, we marginalize over $\phi_R$. This integration is approximated using Gauss-Hermite quadrature, resulting in an approximate expression for the pdf of $X_{1,1}$ as
\vspace{-2mm}
    \begin{align}
    \hspace{-2.5mm}
        f_{X_{1,1}}(x) \approx \pi^{-3/2}\sum_{i=1}^N w_i\big[(x\hspace{-0.5mm}-\hspace{-0.5mm}A_{1,i})(A_{1,i}\hspace{-0.5mm}+\hspace{-0.5mm}B_{1,i}\hspace{-0.5mm}-\hspace{-0.5mm}x)\big]^{-\frac{1}{2}},  \tag{24}
    \end{align}
where $A_{1,i} = a_{1,1}\text{c}^2(\phi_{1,i})$, $B_{1,i} = a_{1,2}\text{s}^2(\phi_{1,i})$, and $\phi_{1,i} = \sqrt{2}\sigma_{\phi_R}u_i+\mu_{\phi_R}$. Here, $u_i$ and $w_i$ are the nodes and weights of the Gauss–Hermite polynomial, respectively. Applying the theorem for the product of two independent RVs, $f_{Y_1}(z) = \int_{-\infty}^{\infty}\frac{1}{|x|}f_{Y_{1,1}}\left(\frac{z}{x}\right)f_{X_{1,1}}(x)dx$, the pdf of $Y_1$ is approximated as
\vspace{-4mm}
    \begin{align}\label{equ: f_Y1_1}
        &f_{Y_1}(z) \approx \pi^{-3/2}\sum_{k=1}^3\hspace{-0.5mm}\sum_{i=1}^N a_kw_iz^{-b_k}I_{k,i}(z),\tag{25}\\
        &I_{k,i}(z) \hspace{-0.5mm}= \hspace{-0.5mm}\int_{A_{1,i}}^{A_{1,i}+B_{1,i}}\hspace{-2.5mm}x^{b_k-1} \big[(x-A_{1,i})(A_{1,i}\hspace{-0.5mm}+\hspace{-0.5mm}B_{1,i}-x)\big]^{-\frac{1}{2}}dx.\nonumber
    \end{align}
To evaluate $I_{k,i}(z)$, we apply the variable transformation, $x = A_{1,i}+B_{1,i}\text{s}^2\theta$, and rearrange the expression into the standard form of the Gauss hypergeometric function. Utilizing the integral identity $\int_{0}^1u^{c-1}(1-u)^{d-1}(1-ku)^{-a}du = B(c,d)_2F_1(a,c;c+d;k)$, where $B(c,d)$ is the Beta function, and $_2F_1(.)$ is the hypergeometric function, $I_{k,i}(z)$ can be evaluated in closed-form. Substituting the results into \eqref{equ: f_Y1_1}, the pdf $f_{Y_1}(z)$ can be approximated as 
\vspace{-2mm}
\begin{align}\label{equ: f_Y1_2}
        f_{Y_1}(z) \hspace{-0.5mm}\approx \hspace{-0.5mm}{\pi}^{-\frac{1}{2}}\hspace{-1.5mm}\sum_{k=1}^3\hspace{-1.0mm}\sum_{i=1}^N a_kw_iz^{-b_k}\hspace{-0.5mm} A_{1,i}^{b_k-1}\hspace{-0.5mm}_2F_1\hspace{-0.5mm}\left(\hspace{-0.5mm}1\hspace{-0.5mm}-\hspace{-0.5mm}b_k,\hspace{-0.5mm}\frac{1}{2};1;-\frac{B_{1,i}}{A_{1,i}}\hspace{-0.5mm}\right)\hspace{-0.5mm}.\tag{26}
        \vspace{-5mm}
    \end{align}
The steps outlined previously can be similarly applied to the  RV $Y_2 = (r^2+h^2)^{-(m+3)}(a_{2,1}\text{c}^2\phi_R+a_{2,2}\text{s}^2\phi_R\text{c}^2(\theta-\theta_R))$. Once, the pdfs of $Y_1$ and $Y_2$ are obtained, the pdf of their sum, $Y_{1+2} = Y_1+Y_2$ can be approximated using the convolution theorem as
\vspace{-2mm}
\begin{align}
        \hspace{-2.6mm}f_{Y_{1+2}}(z) \hspace{-0.5mm}\approx \hspace{-0.5mm}\frac{1}{\pi}\hspace{-1.5mm}\sum_{k,i}\sum_{k',i'} \hspace{-0.5mm}C_{k,i}C'_{k',i'}z^{1-b_k-b'_{k'}}B(1\hspace{-0.5mm}-\hspace{-0.5mm}b_k,1\hspace{-0.5mm}-\hspace{-0.5mm}b'_{k'}), \tag{27}
        \vspace{-7mm}
    \end{align}
where $C_{k,i}=a_kw_iA_{1,i}^{b_k-1}{}_2F_1\hspace{-0.5mm}\left(\hspace{-0.5mm}1\hspace{-0.5mm}-\hspace{-0.5mm}b_k,\frac{1}{2};1;-\frac{B_{1,i}}{A_{1,i}}\right)$. The term $Y_3 = \sum_{k=0}^{\infty}a_{k,3}\binom{\frac{1}{2}}{k}r^{2k+1}(r^2+h^2)^{(k-m-\frac{5}{2})}\text{c}(\theta-\theta_R)\text{s}(2\phi_R)$ is observed to be a sum of numerous small random variables. Therefore, by invoking the central limit theorem, $Y_3$ can approximated as a zero-mean Gaussian RV i.e., $Y_3\approx \mathcal{N}(0,\sigma_{Y_3}^2)$, where the variance is given by $\sigma_{Y_3}^2 = \frac{1-\text{exp}(-8\sigma_{\phi_R}^2)\text{c}(4\mu_{\phi_R})}{4}\hspace{-1.5mm}\sum_{k=0}^{\infty}a_{k,3}^2\binom{\frac{1}{2}}{k}^2\hspace{-1.5mm}\sum_{j=1}^{3}\hspace{-0.5mm}a_j\hspace{-0.5mm}h^{\alpha+\beta+1}B_{u_1}^{u_2}(\alpha/2+1,-\alpha/2-\beta)$, with $\alpha= 4k+z+b_j$, $\beta = 2k-2m-5$, and $u_i = y_i^2/(y_i^2+h_i^2)$. Here, $B_{u_1}^{u_2}(.)$ denotes the incomplete Beta function. Finally, using the convolution of $Y_{1+2}$, and $Y_3$, along with Laplace transform techniques, the pdf of the SNR under the VULO scheme can be derived in closed-form as shown in~\eqref{equ:out_VULO_111}, where $G_{1,0}^{0,1}(.)$ is the Meijer function. The corresponding outage probability can then be evaluated by integrating the SNR pdf up to a threshold $\gamma_{th}$. Employing the Gauss quadrature over the interval $(0,\gamma_{th})$, the resulting approximation for the outage probability is presented in~\eqref{equ:out_VULO_112}.

    \vspace{-1mm}
\section{Numerical Results and Discussion}
\vspace{-1mm}
We present numerical results to verify our outage probability analysis and illustrate the performance gains of the TLL-assisted VLC systems. Unless stated explicitly, in all simulations, we have set the parameter values to, $\mu_{\phi_R} = 20^{\circ}$, $\sigma_{\phi_R}^2 = 8^{\circ}$, $A=10^{-4}$ m$^{2}$, $\Phi_{FoV} = 90^\circ$, $\theta_{1/2}=30^\circ$, $R=0.75$~A/W, $\gamma_{th} = 5$, $P_S = 8$ dBW, $R_{circ} = 5$ m, $\sigma^2 = 10^{-12}$, and $n_l = 1.33$.

\begin{figure}
    \centering
    \includegraphics[width=0.8\linewidth]{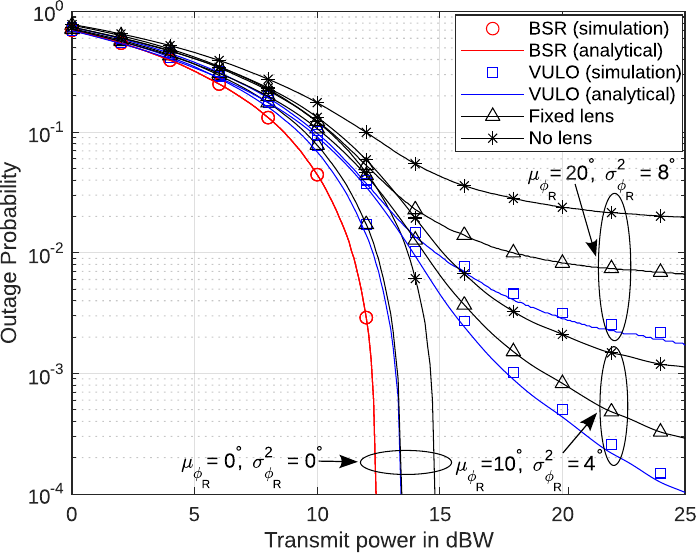}
    \vspace{-3mm}
    \caption{Outage probability vs. transmit power at the AP, $P_S$ in dBW.}
    \vspace{-5mm}
    \label{fig:out_p}
\end{figure}

To demonstrate the performance gains of the BSR and VULO schemes, we compare them against two conventional VLC receivers: with a fixed lens and a receiver without a lens.
\subsubsection{Fixed Lens} In this scenario, we assume that the lens is fixed, and no orientation adjustment is possible, hence, $\psi_x = 0^{\circ}$ and $\psi_y= 0^{\circ}$. Moreover, $\hat{\boldsymbol{\eta}}_{len}\hspace{-1.5mm}\cdot\hat{\boldsymbol{\eta}}_{R} = 1$, $\text{c}(\alpha_{L,R}) = \hat{\boldsymbol{\eta}}_{TR}\cdot \hat{\boldsymbol{\eta}}_{R}$, and hence, $\text{c}(\phi_{L,R})$ can be simplified to
    \vspace{-2.5mm}
    \begin{align}\label{equ_123}
        \text{c}\phi_{L,R} = n_l^{-1}((n_l^2-1)+(\hat{\boldsymbol{\eta}}_{TR}\cdot \hat{\boldsymbol{\eta}}_{R})^2)^{\frac{1}{2}}. \tag{30}
        \vspace{-2mm}
    \end{align}

\subsubsection{No Lens} This scenario represents a typical VLC receiver without a lens. When no lens is deployed at the receiver, $\text{c}\phi_{L,R}$ corresponds to the angle between the transmitter-receiver trajectory and the receiver's axis. Consequently, $\text{c}(\phi_{L,R})$ can be expressed as 
    \vspace{-2.5mm}
    \begin{align}
        \text{c}(\phi_{L,R}) = \hat{\boldsymbol{\eta}}_{TR}\cdot\hat{\boldsymbol{\eta}}_{R}. \tag{31}
        \vspace{-2mm}
    \end{align}
    
Fig. \ref{fig:out_p} illustrates the outage probability versus the transmit power at the AP, expressed in dBW. The results are presented for the BSR and VULO schemes and are compared with scenarios employing a fixed lens and a receiver without any lens. The analytical expressions are compared with simulation results, where the channel gain is computed using (3), confirming the accuracy of the derived expressions. It is observed that the proposed TLL-assisted VLC receiver significantly enhances the outage performance compared to both the fixed lens and conventional receivers, under a wide range of random receiver orientation conditions.  Among the schemes, the BSR scheme achieves the best performance, and is independent of random receiver orientation condition, while the VULO scheme also offers notable improvements with lower complexity. Specifically, the outage probability is improved from $1\times 10^{-1}$ to 
$3\times 10^{-3}$ at a transmit power of 
$12$ dBW under a $8^{\circ}$ polar angle variation in random receiver orientation using the BSR scheme. Except for the BSR scheme, the outage probability in all other schemes saturates at a high $P_S$ values when $\sigma_{\phi_R}^2>0$, due to unavoidable misalignment at large $\phi_R$ angles. It is noted that performance improvements achieved by both the BSR and VULO schemes become more significant at higher $\sigma_{\phi_R}^2$ values, as they effectively mitigate the losses caused by orientation errors. In the absence of random receiver orientation, the performance of the VULO scheme coincides with that of the fixed lens case, since in both schemes the liquid surface and the receiver surface remain horizontally aligned.

\begin{figure}
\vspace{-3mm}
    \centering
    \hspace{3mm}\includegraphics[width=0.9\linewidth]{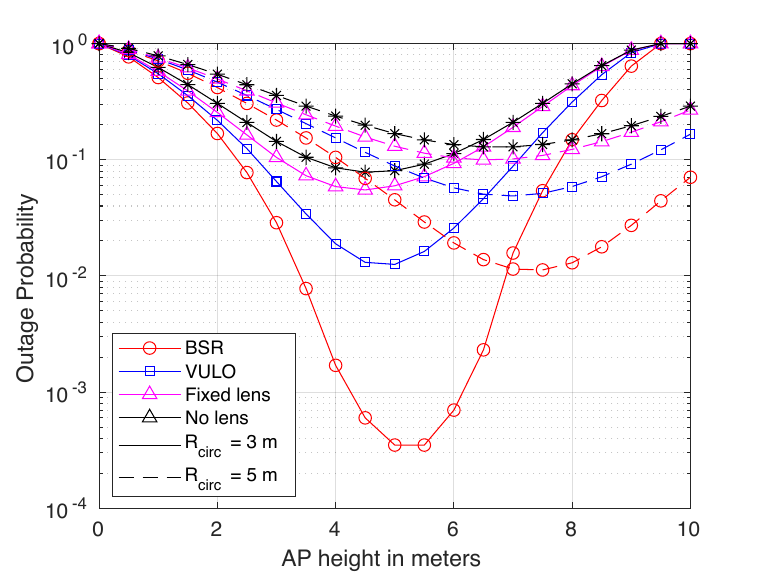}
    \vspace{-3mm}
    \caption{Outage probability vs. height of the AP in meters.}
    \label{fig:out_h}
    \vspace{-8mm}
\end{figure}

Fig. \ref{fig:out_h} presents the outage probability versus the AP height in meters. Results are provided for the proposed schemes and benchmark approaches under different radii of the RWP model, $R_{circ}$. It is observed that an optimal height exists for each scheme under a given setting, which is expected due to the directive nature of the LED lighting patterns. At smaller height values values, the coverage area on the floor is small whereas at larger height values, the channel becomes weaker. The optimal height value for the BSR scheme is slightly higher than that for the VULO and other benchmark schemes. Moreover, the optimal height value increases with the $R_{circ}$ value. In particular, the optimal $h$ is approximately $5.5$ m and $7.5$ m for $R_{\circ} = 3$ m and $R_{\circ} = 5$ m, respectively. At lower $R_{circ}$ values, user tends to move closer to the LED, resulting in lower outage probabilities. Consequently, the minimum achievable outage probability increases with $R_{circ}$. 

\vspace{0mm}
\section{Conclusion}
\vspace{0mm}
In this paper, we investigated an electrowetting surface-based cuboid TLL-assisted VLC system, considering practical factors such as user mobility and random receiver orientation. A precise channel gain model suitable for the TLL-assisted VLC system was developed. To enhance signal reception, we optimized the orientation angles of the TLL, proposing two adaptive lens orientation schemes, namely ($i$) the BSR scheme and ($ii$) the VULO scheme. We derived closed-form analytical expressions for the outage probability of each scheme. Numerical results demonstrate that the proposed TLL-assisted receiver and orientation control schemes significantly outperform the benchmarks in terms of outage performance. Specifically, the BSR scheme achieves an improvement in outage probability from $1\times 10^{-1}$ to $3\times 10^{-3}$ at a transmit power of $12$ dBW and under a $8^{\circ}$ polar angle variation in random receiver orientation.

\vspace{-1mm}
\bibliographystyle{IEEEtran}
\bibliography{Main}

\end{document}